\documentclass[twocolumn]{jpsj3}
\usepackage{txfonts}
\usepackage{braket}

\title{Self-Consistent Fluctuation Theory for Strongly Correlated Electron Systems}

\def\runauthor{H. \textsc{Kusunose}}
\author{Hiroaki \textsc{Kusunose}\thanks{kusu@phys.sci.ehime-u.ac.jp}}

\inst{Department of Physics, Ehime University, Matsuyama, Ehime 790-8577}

\abst{
A self-consistent theory for two-particle fluctuations with renormalized irreducible vertices is proposed.
Using the Parquet formalism, we construct the fully antisymmetric full vertex in terms of the two-particle fluctuations in the charge, the spin and the particle-particle channels on an equal footing to satisfy the Pauli principle.
The fluctuations are determined self-consistently, which are reflected into the one-particle self-energy via the Schwinger-Dyson equation.
We demonstrate the application of the present theory to the impurity Anderson model and the Hubbard model on a square lattice mainly for the particle-hole symmetric case.
In both models the vertex renormalization in the spin channel eliminates magnetic instabilities of mean-field theory to ensure the Mermin-Wagner theorem.
The present theory gives the same critical exponents of the self-consistent renormalization theory in the quantum critical region.
}

\kword{two-particle self-consistency, parquet equation, self-consistent renormalization theory, impurity Anderson model, Hubbard model, Mermin-Wagner theorem, quantum critical point}

\begin{document}
\maketitle

\section{Introduction}
A variety of phenomena such as a metal-insulator transition\cite{Mott49,Mott90,Imada98,Castellani79,Kotliar00,Limelette03} and an anisotropic superconductivity\cite{Onuki07,Moriya00,Yanase03,Manske04} emerges by strong electron-electron correlations.
The fundamental models like the Hubbard and the Anderson lattice models have been investigated extensively using various analytical and numerical techniques.
It has revealed that microscopic detail of the non-interacting band structures is capable to yield various type of metallic states and superconducting pairings
in spite of simplicity of the models.
Insulating phases with or without a magnetic ordering are described by the models as well.

The attempts to describe effects of strong correlations in metallic phases often begin with the Hartree-Fock (HF) and the random-phase approximation (RPA).
Then, the low-order perturbation theory and the fluctuation exchange (FLEX) approximation\cite{Bickers89,Bickers89a,Bickers91} that includes systematically the collective fluctuations improve the treatment of correlations.
They are mainly based on the approximation for the one-particle self-energy, and the latter follows the Baym and Kadanoff's conserving arguments\cite{Baym61,Baym62} within the one-particle level\cite{conservingapprox}.
So-called the dynamical mean-field theory (DMFT) taking account of local fluctuations exactly\cite{Georges96,Georges04,Maier05} is also classified into the one-particle based theories.
The theories mentioned above have achieved considerable success in some aspects, however, at the same time they have significant inconsistency in the two-particle level such as a violation of the Mermin-Wagner theorem\cite{Mermin66,flexmw}.
This is inherent from the one-particle based theories and/or the lack of (spatial) vertex corrections\cite{dmftvc,Katanin09}.
These theories also violate the Pauli principle that is the essential ingredient behind the Fermi-liquid (FL) theory\cite{Shankar94,Metzner98,Chitov95,Chitov98}.
We also note that an appropriate description for a magnetic instability especially near a quantum critical point (QCP) facilitates a better understanding of anisotropic superconductivity mostly appeared near the QCP\cite{Onuki07,Moriya00}.

Meanwhile, the theories based on the two-particle self-consistency have been developed.
The successful phenomenological method is the self-consistent renormalization (SCR) theory by Moriya and co-workers\cite{Moriya00,Moriya85,Moriya95,Kawabata74,Moriya00}.
The SCR is thermodynamically consistent due to the self-consistent treatment of spin fluctuations in the mode-mode coupling term, and it gives the non-trivial exponents near the QCP.
Since the SCR theory relies on the phenomenological expression for the spin fluctuation that is valid in the FL regime, a range of the applicability of the SCR is a delicate issue especially in heavy-fermion systems where a crossover to a local-moment regime occurs at elevated temperatures.
Note that in the quantum critical region, the renormalization-group treatments give the same exponents of the SCR\cite{Hertz76,Millis93}.
The similar idea of the SCR has been used in the microscopic theory referred as the two-particle self-consistent (TPSC)\cite{Vilk97,Senechal04}, where the charge and the spin fluctuations in the longitudinal channel are determined by enforcing the local sum rules to satisfy the Pauli principle.
The double occupancy in the theory is expressed empirically in terms of the renormalized irreducible vertex used to describe the spin fluctuation, which introduces the self-consistent relation of the fluctuations.
Once the two-particle fluctuations are obtained, the single-particle self-energy is computed by the way similar to the paramagnon theories.

In order to go beyond the FLEX, Bickers and the collaborators have argued the full consistency among one and two-particle levels using the parquet formalism\cite{Bickers91,Senechal04,Dominicis64,Chen92}.
Note that the parquet formalism with the unrenormalized bare irreducible vertices is equivalent to the FLEX up to the two-particle level\cite{alprocess}.
They have approximated the irreducible vertices as the contact-type pseudo potentials at the representative point in the momentum-frequency space in order to solve the parquet equation practically.
The quantitative agreements have been demonstrated by the comparison with the quantum monte carlo (QMC) data for the intermediate repulsions.
Nevertheless, there exists a methodological difficulty in choosing the representative point in the 4-vector space for general models, which is usually unknown a priori.
For the impurity Anderson model Jani\v{s} and the co-worker have found the semi-analytic solution for the simplified parquet equation in the critical regime of the spin fluctuations\cite{Janis07,Janis08}.
They have reproduced the correct Kondo scale in the strong-coupling limit.

These findings strongly encourage to construct a reasonable microscopic theory for strongly correlated electron systems using the parquet formalism with appropriate modifications to the previous attempts.
The parquet formalism is also suitable for the purpose that the two-particle self-consistency should be enforced prior to the single-particle one in view of the loop-expansion argument of the renormalization-group theory.
Furthermore, the parquet equation is to be thermodynamically consistent, since it can be derived by functional derivatives of the generating free-energy functional with respect to the two-particle vertex functions\cite{Janis98}.
It is not a conserving approximation in the sense of Baym and Kadanoff\cite{Bickers91,Senechal04}.
Needless to say, conserving approximations do not necessarily mean quantitatively reliable approximations.
Indeed, the FLEX shows the considerable deviations from the finite-size QMC results even in the weak-coupling region\cite{Bickers91,Vilk97}.
Moreover, one should notice that any expression for the self-energy following the Baym-Kadanoff scheme is inconsistent with the exact self-energy formula of the Schwinger-Dyson equation\cite{Bickers91,Vilk97,Janis98}.
The both relations could be satisfied simultaneously, only when the exact expression for the irreducible vertices would be found\cite{Takada95}.

In this paper, we propose yet another self-consistent fluctuation (SCF) theory based on the parquet formalism, which will be explained in the next section.
The formulation of the SCF theory will be given in \S3, where the fully antisymmetric full vertex is constructed in terms of the two-particle fluctuations in all channels to ensure the Pauli principle.
Regarding the Pauli principle, the present theory is contrast to the TPSC, which satisfies it implicitly through the local sum rules.
These fluctuations are determined self-consistently, and they are used to improve the one-particle self-energy in the same spirit of the TPSC.
Note that the numerical cost of the present theory is much less than that of the FLEX.
In \S4, we demonstrate the applications to the impurity Anderson model and the Hubbard model on a square lattice as a benchmark of the present theory.
In the last section, the relation to the SCR theory in the QCP region is discussed\cite{Bickers92}.
The present theory gives the same critical exponents and it can describe a crossover to non-universal region away from the QCP without introducing any phenomenological cutoffs as in the SCR.
We also argue the possible improvements of the present theory, and we summarize the paper.
In Appendix we summarize the derivation of the parquet equation for the single-band model in the case of a nonlocal general two-body interaction conserving the $z$-component of the spins.

\section{Parquet Formalism}

In this section, we introduce briefly the parquet formalism and set up the notational convention through it.
A detailed pedagogical review for the parquet formalism can be found in the literature\cite{Senechal04}.

\subsection{Full vertex and crossing symmetry}
\begin{figure}[tb]
\begin{center}
\includegraphics[width=6.5cm]{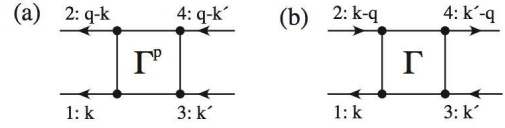}
\end{center}
\caption{The full vertices in (a) the PP channel and (b) the PH channel. The four-vector momenta are assigned to each legs as shown in the figure. The PP and the PH pairs carry the momentum $q$ from right to left.}
\label{full_vertex}
\end{figure}

Let us first define an effective interaction with full vertex in the particle-particle (PP) and the particle-hole (PH) channels in the form,
\begin{align}
&\text{PP channel}=\frac{1}{4}\Gamma^{p}(12;34)c^{\dagger}(1)c^{\dagger}(2)c^{}(4)c^{}(3),
\cr
&\text{PH channel}=\frac{1}{4}\Gamma(12;34)c^{\dagger}(1)c^{}(2)c^{\dagger}(4)c^{}(3),
\label{effint}
\end{align}
where the number in the parenthesis represents a set of indices of the band (orbital), the spin, the spatial coordinate and the imaginary time.
The superscript ``p'' indicates a quantity concerning the PP channel.
The repeated indices that do not appear in the left-hand side are assumed to be summed hereafter.
We define the fourier and inverse fourier transformations between the spatial coordinates and the four-vector momenta $k=(\mib{k},i\omega_{n})$, $k'=(\mib{k}',i\omega_{n}')$, $q=(\mib{q},i\epsilon_{m})$ as
\begin{align}
&A(12;34)=\sum_{kk'q}A(k,k';q)e^{i\left[k.1-(k-q).2-k'.3+(k'-q).4\right]},
\cr
&A(k,k';q)=\sum_{234}A(12;34)e^{-i\left[k.1-(k-q).2-k'.3+(k'-q).4\right]},
\end{align}
where $\omega_{n}$, $\omega_{n}'$ are the fermionic Matsubara frequencies, while $\epsilon_{m}$ is the bosonic one.
We have assigned the four momenta at each legs of the vertex as $A(k,k';q)\equiv A(k,k-q;k',k'-q)$ in the PH channel and $A^{p}(k,k';q)\equiv A^{p}(k,q-k;k',q-k')$ in the PP channel.
The assignment of the variables for the full vertices are schematically shown in Fig.~\ref{full_vertex}.

The antisymmetric nature of the fermion fields gives the following relations for the full vertices,
\begin{align}
&\Gamma^{p}(D)=-\Gamma^{p}(T),
\quad
\Gamma(D)=-\Gamma(C),
\cr
&\Gamma^{p}(D)=-\Gamma(P)=\Gamma(X),
\label{cs2}
\end{align}
where we have introduced sets of coordinates/momenta as
\begin{align}
&D\equiv(12;34),\,\,(k,k';q),
\cr
&T\equiv(12;43),\,\,(k,q-k';q),
\cr
&C\equiv(13;24),\,\,(k,k-q;k-k'),
\cr
&P\equiv(14;32),\,\,(k,k';k+k'-q),
\cr
&X\equiv(13;42),\,\,(k,q-k';k-k'),
\end{align}
for notational convenience.
These relations are called the crossing symmetry, which is the direct consequence of the Pauli principle.
Note that frequently used theories such as the HF-RPA and FLEX violate the crossing symmetry.

\subsection{Bethe-Salpeter equation}
The full vertex can be decomposed into multiple scatterings of the PP or the PH pair, respectively, in terms of the irreducible vertices, $\Gamma^{0p}(D)$ and $\Gamma^{0}(D)$.
Namely, so-called the Bethe-Salpeter (BS) equation in the PH channel is given by
\begin{equation}
\Gamma(D)=\Gamma^{0}(D)+\Phi(D),
\quad
\Phi(D)\equiv-\left[\Gamma^{0}\chi_{0}\Gamma\right](D),
\end{equation}
where $\chi_{0}(12;34)=-G(13)G(42)$ is the lowest-order two-particle green's function.
$\Gamma^{0}(12;34)=\Gamma^{0}(43;21)$, is irreducible with respect to a vertical cut of two green's function lines.
Here the ``matrix product'' is introduced as
\begin{equation}
\left[AB\right](D)=A(12;56)B(56;34),
\,\,\sum_{p}A(k,p;q)B(p,k';q).
\end{equation}
The identity matrix is given by $I(D)=\delta(13)\delta(24),\,\,\delta_{kk'}$.

\begin{figure}[tb]
\begin{center}
\includegraphics[width=8.5cm]{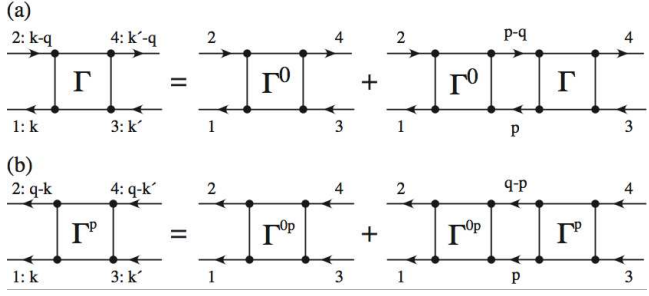}
\end{center}
\caption{The Bethe-Salpeter equations in (a) the vertical PH channel, and (b) the PP channel. $\Gamma^{0}$ and $\Gamma^{0p}$ are irreducible with respect to a vertical cut of two green's function lines.}
\label{bs}
\end{figure}

Similarly, the BS equation in the PP channel is given by
\begin{equation}
\Gamma^{p}(D)=\Gamma^{0p}(D)+\Psi(D),
\quad
\Psi(D)\equiv-\frac{1}{2}\left[\Gamma^{0p}\psi_{0}\Gamma^{p}\right](D),
\end{equation}
where $\psi_{0}(12;34)=G(13)G(24)$.
The factor $1/2$ is to avoid double counting of twisted diagrams.
Exchanging the indices $1$ and $2$ in the BS equation, we can show that $\Gamma^{0p}$ also satisfies the crossing symmetry (\ref{cs2}), {\it i.e.}, $\Gamma^{0p}(D)=-\Gamma^{0p}(T)$.
Note that the irreducible vertex in one channel is reducible in another in general.
The Feynman diagrams for the BS equations are given in Fig.~\ref{bs}.

Using the full vertices, we express two-particle green's functions in the PP and the PH channels as
\begin{align}
\psi(D)&\equiv\Braket{T_{\tau}c^{}(1)c^{}(2)c^{\dagger}(4)c^{\dagger}(3)}
\cr
&=\psi_{0}(D)-\frac{1}{2}\left[\psi_{0}\Gamma^{0p}\psi\right](D)
=\psi_{0}(D)-\frac{1}{2}\left[\psi_{0}\Gamma^{p}\psi_{0}\right](D),
\cr
\chi(D)&\equiv\Braket{T_{\tau}c^{}(1)c^{\dagger}(2)c^{}(4)c^{\dagger}(3)}
\cr
&=\chi_{0}(D)-\left[\chi_{0}\Gamma^{0}\chi\right](D)
=\chi_{0}(D)-\left[\chi_{0}\Gamma\chi_{0}\right](D).
\end{align}
The ordinary two-particle correlation functions are obtained by setting $1=2$ and $3=4$, {\it i.e.}, $\chi(q)=\sum_{kk'}\chi(k,k';q)$ and $\psi(q)=\sum_{kk'}\psi(k,k';q)$, where $\sum_{k}\equiv\frac{T}{N_{0}}\sum_{\mib{k}}\sum_{n}$ with $N_{0}$ being the number of lattice sites.

\subsection{Parquet equation}
Although the equations given in the previous subsections are formally exact, it is necessary to introduce physically motivated approximations to make them practically solvable.
However, a simple approximation to the irreducible vertices often results in full vertices violating the crossing symmetry.
In order to obtain a solution that maintains the crossing symmetry, it is adequate to use a set of exact relations between the irreducible vertices.
This is the Parquet equation\cite{Senechal04,Dominicis64}, which is give by
\begin{align}
&\Gamma^0(D)=\gamma(D)-\Phi(C)-\Psi(P),
\cr
&\Gamma^{0p}(D)=\gamma^p(D)+\Phi(X)-\Phi(P).
\label{parqueteq}
\end{align}
Here we have introduced the fully irreducible vertex, $\gamma(D)$, in all channels, which satisfies the crossing symmetry like (\ref{cs2}).
A choice of $\gamma(D)$ is ambiguous.
In the simplest case, an anti-symmetrized bare interaction $U(D)$ is used for $\gamma(D)$, and it is often called as ``basic'' parquet equation.

\begin{figure}[tb]
\begin{center}
\includegraphics[width=8.5cm]{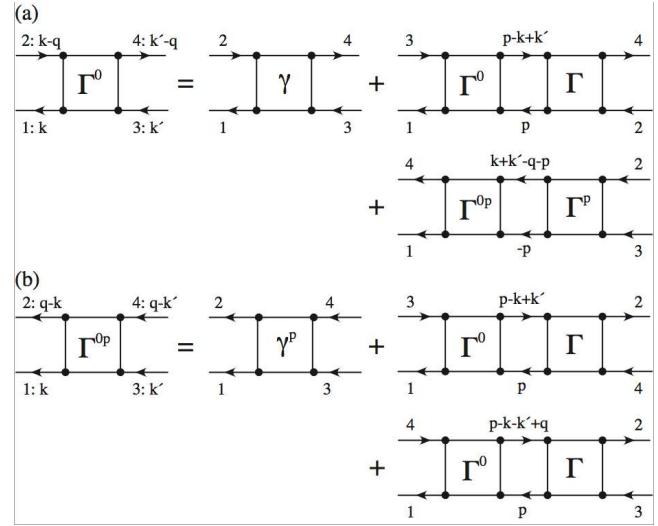}
\end{center}
\caption{The Parquet equations for the irreducible vertices in (a) the PH channel, and (b) the PP channel. By the BS equation with these irreducible vertices, the resultant full vertices satisfy the crossing symmetry.}
\label{parquet}
\end{figure}

Using the BS equations, we reexpress the full vertices as
\begin{align}
&\Gamma(D)=\gamma(D)+\Phi(D)-\Phi(C)-\Psi(P),
\cr
&\Gamma^{p}(D)=\gamma^{p}(D)+\Psi(D)+\Phi(X)-\Phi(P),
\end{align}
where the crossing symmetry is apparent.
It should be emphasized that resultant full vertices always satisfy the crossing symmetry, irrespective of the explicit expressions of $\Phi$ and $\Psi$.
The Feynman diagrams of the parquet equations are shown in Fig.~\ref{parquet}.

The BS and the parquet equations constitute a set of the self-consistent integral equations among the full and the irreducible vertices for an arbitrary given one-particle green's function $G(k)$.

\subsection{Self energy in terms of full vertex}
\begin{figure}[tb]
\begin{center}
\includegraphics[width=8.5cm]{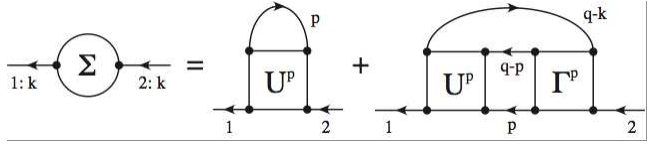}
\end{center}
\caption{The Schwinger-Dyson equation for the self-energy in terms of the full vertex.}
\label{se}
\end{figure}

Once we obtain the full vertex, the self-energy is calculated through the Schwinger-Dyson equation as shown schematically in Fig.~\ref{se},
\begin{align}
&\Sigma(12)=\Sigma_{\rm HF}(12)+\widetilde{\Sigma}(12),
\cr
&\quad \Sigma_{\rm HF}(12)=U^{p}(13;24)G(43),
\cr
&\quad \widetilde{\Sigma}(12)=-\frac{1}{2}\left[U^{p}\psi_{0}\Gamma^{p}\right](13;24)G(43).
\end{align}
This equation together with the BS and the parquet equations provide a complete set of the self-consistent equations for correlated fermion systems.

\subsection{Case of Hubbard-type interaction with SU(2) symmetry}
In the presence of SU(2) symmetry in spin space, the two-particle quantities can be classified by the total spin of the PH and the PP pairs, and the one-particle quantities are independent of spin indices (See, Appendix in detail).
Namely, the charge (even-parity) and the spin (odd-parity) vertices belong to $S=0$ and $S=1$ PH (PP) channels, respectively.
For later convenience, we quote the relevant expressions in the case of the Hubbard-type bare interaction, $v(12)=U\delta(12)$ with SU(2) symmetry.

\vspace{2mm}\noindent\underline{BS equation}
\begin{align}
&\Gamma_{\xi}(D)=\Gamma_{\xi}^{0}(D)+\Phi_{\xi}(D),
\quad
(\xi=c,s),
\cr&\quad\quad
\Phi_{\xi}(D)=-\sum_{p}\Gamma_{\xi}^{0}(k,p;q)\chi_{0}(p;q)\Gamma_{\xi}(p,k';q),
\cr
&\Gamma_{\zeta}(D)=\Gamma_{\zeta}^{0}(D)+\Psi_{\zeta}(D),
\quad
(\zeta=e,o),
\cr&\quad\quad
\Psi_{\zeta}(D)=-\frac{1}{2}\Gamma^{0}_{\zeta}(k,p;q)\psi_{0}(p;q)\Gamma_{\zeta}(p,k';q),
\label{bssu2}
\end{align}
where $\chi_{0}(k;q)=-G(k)G(k-q)$ and $\psi_{0}(k;q)=G(k)G(q-k)$.

\vspace{2mm}\noindent\underline{Two-particle correlation function}
\begin{align}
\chi_{\xi}(q)
&=\chi_{0}(q)-\sum_{kk'}\chi_{0}(k;q)\Gamma^{0}_{\xi}(k,k';q)\chi_{\xi}(k';q)
\cr
&=\chi_{0}(q)-\sum_{kk'}\chi_{0}(k;q)\Gamma_{\xi}(k,k';q)\chi_{0}(k';q),
\cr
\psi_{\zeta}(q)
&=\psi_{0}(q)-\frac{1}{2}\sum_{kk'}\psi_{0}(k;q)\Gamma^{0}_{\zeta}(k,k';q)\psi_{\zeta}(k';q)
\cr
&=\psi_{0}(q)-\frac{1}{2}\sum_{kk'}\psi_{0}(k;q)\Gamma_{\zeta}(k,k';q)\psi_{0}(k';q),
\label{twopsu2}
\end{align}
where $\chi_{0}(q)=\sum_{k}\chi_{0}(k;q)$ and $\psi_{0}(q)=\sum_{k}\psi_{0}(k;q)$.
Note that the (pair) density operators with $1/\sqrt{2}$ are used in the definition of the correlation functions, {\it i.e.},
\begin{align}
&\rho_{c}=\frac{1}{\sqrt{2}}c_{\alpha}^{\dagger}c_{\alpha}^{},
\quad
\mib{\rho}_{s}=\frac{1}{\sqrt{2}}c_{\alpha}^{\dagger}\mib{\sigma}_{\alpha\beta}c_{\beta}^{},
\cr
&\rho_{e}=\frac{1}{\sqrt{2}}c_{\alpha}^{}(i\sigma^{y})_{\alpha\beta}c_{\beta}^{},
\quad
\mib{\rho}_{o}=\frac{1}{\sqrt{2}}c_{\alpha}^{}(i\sigma^{y}\mib{\sigma})_{\alpha\beta}c_{\beta}^{}.
\end{align}

\vspace{2mm}\noindent\underline{Basic parquet equation ($\gamma=U$)}
\begin{align}
&\Gamma_{c}^{0}(D)=\gamma-\frac{1}{2}\left[\Phi_{c}+3\Phi_{s}\right](C)
+\frac{1}{2}\left[\Psi_{e}-3\Psi_{o}\right](P),
\cr
&\Gamma^{0}_{s}(D)=-\gamma-\frac{1}{2}\left[\Phi_{c}-\Phi_{s}\right](C)
-\frac{1}{2}\left[\Psi_{e}+\Psi_{o}\right](P),
\cr
&\Gamma_{e}^{0}(D)=2\gamma+\frac{1}{2}\left[\Phi_{c}-3\Phi_{s}\right](X)
+\frac{1}{2}\left[\Phi_{c}-3\Phi_{s}\right](P),
\cr
&\Gamma^{0}_{o}(D)=\frac{1}{2}\left[\Phi_{c}+\Phi_{s}\right](X)
-\frac{1}{2}\left[\Phi_{c}+\Phi_{s}\right](P).
\label{parquetsu2}
\end{align}
Here $\gamma$ is used instead of $U$ in order to keep track of the fully irreducible vertex.

\vspace{2mm}\noindent\underline{Crossing symmetry}
\begin{align}
&\Gamma_{e}(D)=\Gamma_{e}(T),
\quad
\Gamma_{o}(D)=-\Gamma_{o}(T),
\cr
&
\Gamma_{c}(D)=-\frac{1}{2}\left[\Gamma_{c}+3\Gamma_{s}\right](C),
\quad
\Gamma_{s}(D)=-\frac{1}{2}\left[\Gamma_{c}-\Gamma_{s}\right](C),
\cr
&
\Gamma_{e}(D)=\frac{1}{2}\left[\Gamma_{c}-3\Gamma_{s}\right](P)=\frac{1}{2}\left[\Gamma_{c}-3\Gamma_{s}\right](X),
\cr
&
\Gamma_{o}(D)=-\frac{1}{2}\left[\Gamma_{c}+\Gamma_{s}\right](P)=\frac{1}{2}\left[\Gamma_{c}+\Gamma_{s}\right](X).
\end{align}

\vspace{2mm}\noindent\underline{Self energy}
\begin{equation}
\Sigma(k)=Un-\frac{U}{2}\sum_{pq}\psi_{0}(p;q)\left[\Gamma_{e}+3\Gamma_{o}\right](p,k;q)G(q-k),
\label{sesu2}
\end{equation}
where $n=\sum_{k}G(k)e^{i\omega_{n}0_{+}}$ is the electron density per spin.

With this preliminary, we will discuss a practical approximation of the parquet formalism in the next section.

\section{Self-Consistent Fluctuation Approximation}
The parquet formalism is a reliable nontrivial renormalization scheme, where two-particle self-consistency and the crossing symmetry are respected.
However, a practical solution is limited to very small systems since the central nonlinear integral equations with three 4-vector variables are too large to handle even at modern numerical computation facilities\cite{Yang09}.
Therefore, a practical simplification preserving essential ingredient of the parquet formalism is necessary.

\begin{figure}[tb]
\begin{center}
\includegraphics[width=5cm]{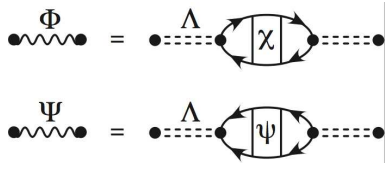}
\end{center}
\caption{The higher-order collective interactions in terms of the two-particle correlation functions.}
\label{fluc}
\end{figure}

In order to simplify the parquet formalism, we introduce the contact-type effective irreducible vertices in the charge, the spin, the even-parity and the odd-parity channels,
\begin{align}
&\Gamma_{c}^{0}(D)\sim \Lambda_{c},
\,\,\,
\Gamma_{s}^{0}(D)\sim -\Lambda_{s},
\cr
&\Gamma_{e}^{0}(D)\sim 2\Lambda_{e},
\,\,\,
\Gamma_{o}^{0}(D)\sim 2\Lambda_{o}\equiv 0,
\label{cvapprox}
\end{align}
which act as the renormalized vertices on an average, and they will be determined later without losing two-particle self-consistency and the crossing symmetry.
The minus sign and the factor 2 are only for the notational simplicity of the following equations.
Note that the effective vertex in the odd-parity channel identically vanishes due to the crossing symmetry, $\Gamma_{o}^{0}(D)=-\Gamma_{o}^{0}(T)$.
With this approximation, the formal solution of the BS equation (\ref{bssu2}) takes the form,
\begin{align}
&\Phi_{\xi}(D)\sim-\Lambda_{\xi}^{2}\chi_{\xi}(q)\equiv\Phi_{\xi}(q),
\quad
(\xi=c,s),
\cr
&\Psi_{e}(D)\sim-2\Lambda_{e}^{2}\psi_{e}(q)\equiv 2\Psi_{e}(q),
\cr
&\Psi_{o}(D)\sim-2\Lambda_{o}^{2}\psi_{o}(q)\equiv0,
\label{phifluc}
\end{align}
which are schematically shown in Fig.~\ref{fluc}.
Here, we have used the first expressions of (\ref{twopsu2}) to yield the two-particle correlation functions,
\begin{align}
&\chi_{c}(q)=\frac{\chi_{0}(q)}{1+\Lambda_{c}\chi_{0}(q)},
\,\,\,
\chi_{s}(q)=\frac{\chi_{0}(q)}{1-\Lambda_{s}\chi_{0}(q)},
\cr
&\psi_{e}(q)=\frac{\psi_{0}(q)}{1+\Lambda_{e}\psi_{0}(q)}.
\label{tprpa}
\end{align}
Using $\Phi_{\xi}(q)$ and $\Psi_{e}(q)$ and the BS equation (\ref{bssu2}), the full vertices from the parquet equation (\ref{parquetsu2}) read,
\begin{align}
&\Gamma_{c}(D)=\gamma+\Phi_{c}(q)-\frac{1}{2}\left[\Phi_{c}+3\Phi_{s}\right](Q)
+\Psi_{e}(\Omega),
\cr
&\Gamma_{s}(D)=-\gamma+\Phi_{s}(q)-\frac{1}{2}\left[\Phi_{c}-\Phi_{s}\right](Q)
-\Psi_{e}(\Omega),
\cr
&\Gamma_{e}(D)=2[\gamma+\Psi_{e}(q)]+\frac{1}{2}\left[\Phi_{c}-3\Phi_{s}\right](Q)
+\frac{1}{2}\left[\Phi_{c}-3\Phi_{s}\right](\Omega),
\cr
&\Gamma_{o}(D)=\frac{1}{2}\left[\Phi_{c}+\Phi_{s}\right](Q)
-\frac{1}{2}\left[\Phi_{c}+\Phi_{s}\right](\Omega),
\label{fvc}
\end{align}
where we have introduced two ``bosonic'' variables, $Q=k-k'$ and $\Omega=k+k'-q$.
The Pauli principle can be confirmed directly by these expressions.

Substituting these vertices into the second definition of the two-particle green's functions (\ref{twopsu2}), we obtain
\begin{align}
&\chi_{c}(q)=\chi_{0}(q)-\chi_{0}(q)^{2}[\gamma+\Phi_{c}(q)]-\Delta\chi_{c}(q),
\cr
&\chi_{s}(q)=\chi_{0}(q)+\chi_{0}(q)^{2}[\gamma-\Phi_{s}(q)]+\Delta\chi_{s}(q),
\cr
&\psi_{e}(q)=\psi_{0}(q)-\psi_{0}(q)^{2}[\gamma+\Psi_{e}(q)]-\Delta\psi_{e}(q),
\cr
&\psi_{o}(q)=\psi_{0}(q),
\label{tpparquet}
\end{align}
where the vertex corrections are defined as
\begin{align}
&\Delta\chi_{c}(q)=\frac{1}{2}\left[\Lambda_{c}^{2}f_{c}(q)+3\Lambda_{s}^{2}f_{s}(q)\right]-\Lambda_{e}^{2}g(q),
\cr
&\Delta\chi_{s}(q)=-\frac{1}{2}\left[\Lambda_{c}^{2}f_{c}(q)-\Lambda_{s}^{2}f_{s}(q)\right]-\Lambda_{e}^{2}g(q),
\cr
&\Delta\psi_{e}(q)=-\frac{1}{2}\left[\Lambda_{c}^{2}h_{c}(q)-3\Lambda_{s}^{2}h_{s}(q)\right].
\label{vc}
\end{align}
Here we have defined the integrals corresponding to the Maki-Thompson processes as
\begin{align}
&f_{\xi}(q)=\sum_{kk'}\chi_{0}(k;q)\chi_{\xi}(k-k')\chi_{0}(k';q),
\cr
&g(q)=\sum_{kk'}\chi_{0}(k;q)\psi_{e}(k+k'-q)\chi_{0}(k';q),
\cr
&h_{\xi}(q)=\sum_{kk'}\psi_{0}(k;q)\chi_{\xi}(k-k')\psi_{0}(k';q).
\end{align}

\begin{figure}[tb]
\begin{center}
\includegraphics[width=7cm]{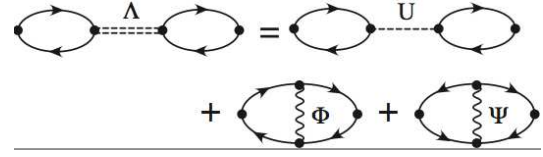}
\end{center}
\caption{The relation between the effective irreducible vertices and the vertex corrections. The $q$-average of this relation gives the self-consistent equation to determine $\Lambda$.}
\label{scf_rel}
\end{figure}

Equating (\ref{tprpa}) and (\ref{tpparquet}), we obtain
\begin{align}
&\left[\Lambda_{\xi}-\gamma\right]\chi_{0}(q)^{2}=\Delta\chi_{\xi}(q),
\cr
&\left[\Lambda_{e}-\gamma\right]\psi_{0}(q)^{2}=\Delta\psi_{e}(q),
\end{align}
which are schematically shown in Fig.~\ref{scf_rel}.
Since we attempt to find a solution within the contact-type vertex approximation, the above equations make sense with an appropriate averaging procedure with respect to $q$.
We adopt a simple average over $q$, and we finally obtain the self-consistent condition for the effective vertices,
\begin{equation}
\Lambda_{\xi}=\gamma+\frac{\sum_{q}\Delta\chi_{\xi}(q)}{\sum_{q}\chi_{0}(q)^{2}},
\quad
\Lambda_{e}=\gamma+\frac{\sum_{q}\Delta\psi_{e}(q)}{\sum_{q}\psi_{0}(q)^{2}}.
\label{effvcond}
\end{equation}
Note that the $q$ average of $f_{\xi}(q)$, $g(q)$ and $h_{\xi}(q)$ appeared in the vertex corrections (\ref{vc}) reduces to the simple average,
\begin{align}
&\sum_{q}f_{\xi}(q)=\sum_{q}h_{\xi}(q)=\sum_{q}\chi_{0}(q)^{2}\chi_{\xi}(q),
\cr
&\sum_{q}g(q)=\sum_{q}\psi_{0}(q)^{2}\psi_{e}(q).
\label{qavvc}
\end{align}
By solving (\ref{vc}), (\ref{effvcond}) and (\ref{qavvc}) self-consistently for given $G(k)$, we obtain the renormalized vertices, $\Lambda_{c}$, $\Lambda_{s}$ and $\Lambda_{e}$, and simultaneously the two-particle correlation functions and the fully antisymmetric vertices via (\ref{fvc}).

\begin{figure}[tb]
\begin{center}
\includegraphics[width=7.5cm]{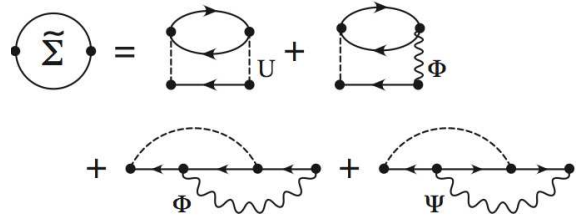}
\end{center}
\caption{The self-energy in the self-consistent fluctuation approximation. Each graphs represent the contributions from the 2nd-order, the longitudinal, the transverse and the PP fluctuations, respectively.}
\label{self}
\end{figure}

Substituting the full vertices (\ref{fvc}) into (\ref{sesu2}) and transforming the variables, we obtain the self-energy
\begin{multline}
\Sigma(k)=Un+U\sum_{q}\biggl[\chi_{0}(q)\left\{\gamma+\frac{1}{2}\left[\Phi_{c}-3\Phi_{s}\right](q)\right\}G(k-q)
\\
-\psi_{0}(q)\Psi_{e}(q)G(q-k)\biggr].
\label{sefluc}
\end{multline}
The self-energy in this approximation is schematically shown in Fig.~\ref{self}.
Note that when we neglect the vertex corrections, $\Delta\chi_c=\Delta\chi_s=\Delta\psi_{e}=0$, the present formalism reduces to the FLEX.

Let us now comment on the self-consistent procedure between two-particle and one-particle levels.
In the two-particle self-consistency, the one-particle propagator is an external input.
It is well known that the one-particle self-consistency without using the consistent frequency-dependent irreducible vertices always results in a universal high-frequency behavior and the Hubbard satellite bands are completely smeared out\cite{Vilk97,Janis07}.
This is because the high-frequency asymptotic behavior sets in, not above the bandwidth but above an incorrect energy scale, $U$, which is extensively discussed by Vilk and Tremblay\cite{Vilk97}, and Jani\v{s} and Augustinsk\'y\cite{Janis07}.

In order to balance the use of the contact-type irreducible vertices, $\Lambda_{\xi}$ and $\Lambda_{e}$, we use the HF propagator as an input for the two-particle self-consistency, and the one-particle self-energy is updated only once.
Specifically, we first determine the HF propagator $G_{0}(k)$ by adjusting the chemical potential to match the given electron density per spin, $n$.
Using $G_{0}(k)$, we determine $\Lambda_{c}$, $\Lambda_{s}$ and $\Lambda_{e}$ by solving the two-particle self-consistent equations.
Then, the self-energy $\Sigma(k)$ is updated via the Schwinger-Dyson equation, and the renormalized chemical potential, $\delta\mu$, is determined\cite{Vilk97} by $\sum_{k}G(k)e^{i\omega_{n}0_{+}}=n$ with $G(k)^{-1}=G_{0}(k)^{-1}-\Sigma(k)+\delta\mu$.

Empirically, the contribution from the PP fluctuation seems to be slightly overestimated, and it yields unphysical behaviors in one-particle spectrum.
By neglecting the last term in (\ref{sefluc}), the SCF works quite well as will be shown in the next section.

\section{Application to Fundamental Models}
In this section, we consider two fundamental models of many-body fermion systems, {\it i.e.}, the impurity Anderson model and the Hubbard model on a square lattice, as applications of the SCF approximation.
In the latter, the first Brillouin zone is discretized by $128\times128$ meshes, and $2048$ Matsubara frequencies are used in both models.
To obtain the real-frequency spectrum, we adopt the Pad\'e interpolation to carry out an analytic continuation.

\subsection{Impurity Anderson model}
The impurity Anderson model is given by
\begin{multline}
H=\sum_{k\sigma}\epsilon_{k}c_{k\sigma}^{\dagger}c_{k\sigma}^{}+\left(E_{f}-\frac{U}{2}\right)\sum_{\sigma}n_{f\sigma}
\\
+\frac{V}{\sqrt{N_{0}}}\sum_{k\sigma}\left(c_{k\sigma}^{\dagger}f_{\sigma}^{}+f_{\sigma}^{\dagger}c_{k\sigma}^{}\right)
+Un_{f\uparrow}n_{f\downarrow},
\end{multline}
where the $f$ level is measured from $-U/2$, and $n_{f\sigma}=f_{\sigma}^{\dagger}f_{\sigma}^{}$.
We use the practically infinite bandwidth of the conduction electron, so that the Hartree green's function is given by
\[
G_{0}(i\omega_{n})^{-1}=i\omega_{n}-E_{f}-U(n_{\rm H}-1/2)+i\Delta_{0},
\]
where $n_{\rm H}=T\sum_{n}G(i\omega_{n})e^{i\omega_{n}0_{+}}$.
The hybridization strength is $\Delta_{0}=\pi \rho_{0}V^{2}$, which is taken as a unit of energy.

Let us consider the symmetric case ($E_{f}=0$), in which $G_{0}(i\omega_{n})$ and $G(i\omega_{n})$ are pure imaginary, and $n=n_{\rm H}=1/2$, $\chi_{0}(i\epsilon_{m})=\psi_{0}(i\epsilon_{m})$ and $\Lambda_{c}=\Lambda_{e}$ hold.

\begin{figure}[tb]
\begin{center}
\includegraphics[width=8.5cm]{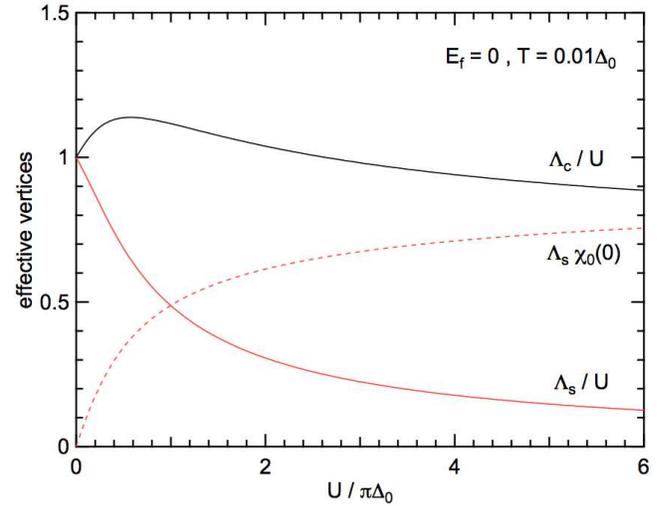}
\end{center}
\caption{(Color online) The $U$ dependence of $\Lambda_{c}$ and $\Lambda_{s}$ for $T/\Delta_{0}=0.01$. Note that the PP vertex $\Lambda_{e}$ is identical with $\Lambda_{s}$ for the symmetric case.}
\label{U_dep}
\end{figure}

The $U$ dependence of the effective irreducible vertices at $T=0.01\Delta_{0}$ are shown in Fig.~\ref{U_dep}.
The charge (even-parity) vertex is roughly the order of $U$, while the spin vertex is strongly renormalized as $U$ increases.
In the HF-RPA theory, the Stoner condition is given by $U/\pi\Delta_{0}=1$, while the spin vertex in the present theory never reaches the magnetic instability condition $\delta\equiv1-\Lambda_{s}\chi_{0}(0)=0$ as it should in the essentially zero-dimensional system.
Note that $\delta$ gives a measure of the Kondo temperature, {\it i.e.}, $\delta\sim \chi_{0}(0)T_{\rm K}$.

\begin{figure}[tb]
\begin{center}
\includegraphics[width=8.5cm]{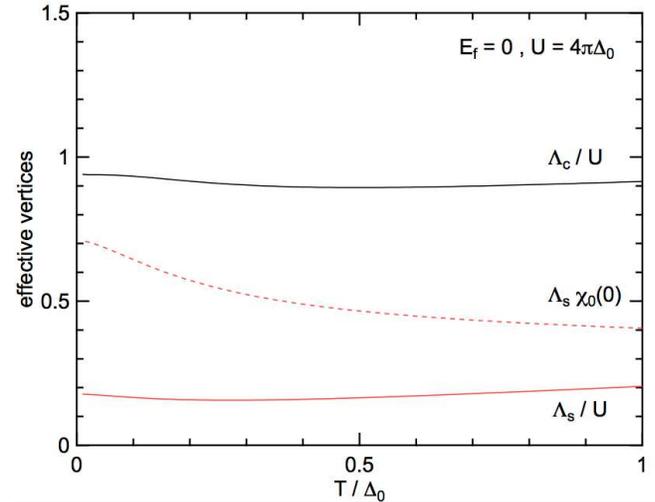}
\end{center}
\caption{(Color online) The $T$ dependence of $\Lambda_{c}$ and $\Lambda_{s}$ at $U/\pi\Delta_{0}=4$.}
\label{T_dep}
\end{figure}

The $T$ dependences of $\Lambda_{c}$ and $\Lambda_{s}$ at $U=4\pi\Delta_{0}$ are shown in Fig.~\ref{T_dep}.
Both the charge and the spin vertices have very weak temperature dependence, and the $T$ dependence of the susceptibilities thus mainly comes from that of $\chi_{0}(i\epsilon_{m})$.

\begin{figure}[tb]
\begin{center}
\includegraphics[width=8.5cm]{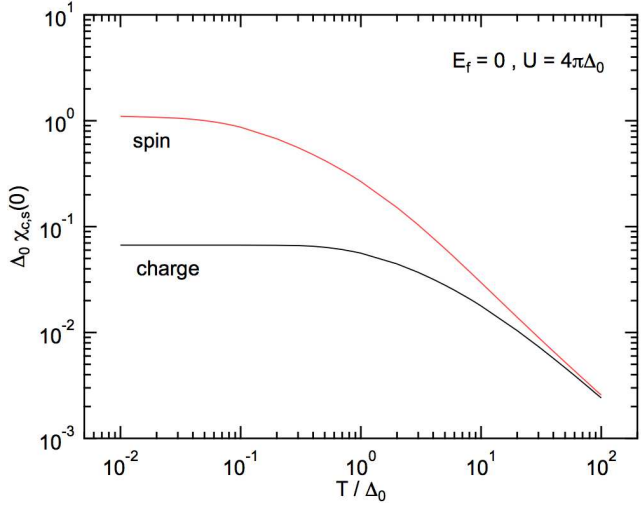}
\end{center}
\caption{(Color online) The $T$ dependence of $\chi_{c}(0)$ and $\chi_{s}(0)$ at $U/\pi\Delta_{0}=4$.}
\label{chi0}
\end{figure}

The $T$ dependence of the static charge and spin susceptibilities, $\chi_{c}(0)$ and $\chi_{s}(0)$ at $U=4\pi\Delta_{0}$ are shown in Fig.~\ref{chi0}.
The charge fluctuation is suppressed below $T\sim U$, while the saturation of the spin susceptibility occurs below $T\sim 0.1\Delta_{0}$ reflecting the singlet formation of the Kondo effect.
The exact expression of the Kondo temperature in Kondo regime\cite{Tsvelik82,Hewson93} ($U, |E_{f}|\to\infty$) is given by
\[
\frac{T_{\rm K}}{\Delta_{0}}=\sqrt{\frac{\pi}{2}\frac{U}{\pi\Delta_{0}}}\exp\left(
-\frac{\pi^{2}}{8}\frac{U}{\pi\Delta_{0}}\left|
1-\left(\frac{2E_{f}}{U}\right)^{2}
\right|
\right).
\]
For the intermediate coupling, $U=4\pi\Delta_{0}$, $T_{\rm K}/\Delta_{0}\simeq 0.018$ and the present theory slightly overestimates the Kondo temperature.

\begin{figure}[tb]
\begin{center}
\includegraphics[width=8.5cm]{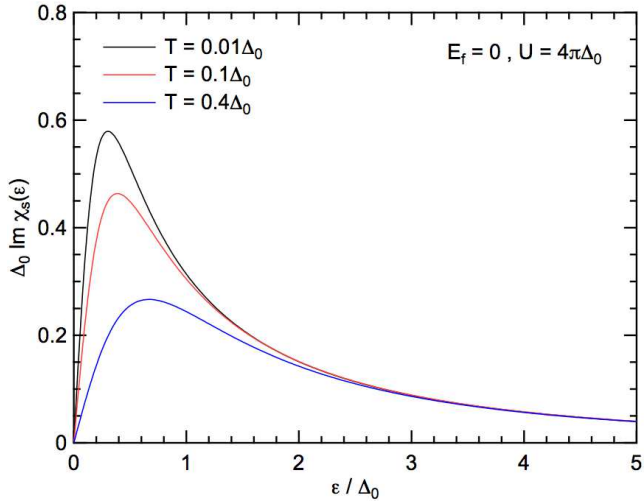}
\end{center}
\caption{(Color online) The $T$ dependence of ${\rm Im}\,\chi_{s}(\epsilon)$ at $U/\pi\Delta_{0}=4$.}
\label{ImChiS_U4}
\end{figure}

The dynamical spin susceptibility is shown in Fig.~\ref{ImChiS_U4}.
As the temperature decreases, the intensity of $\chi_{s}(\epsilon)$ transfers to the lower-energy region.
The low-energy limit of ${\rm Im}\,\chi_{s}(\epsilon)$ is proportional to $\epsilon$ as the FL theory.

\begin{figure}[tb]
\begin{center}
\includegraphics[width=8.5cm]{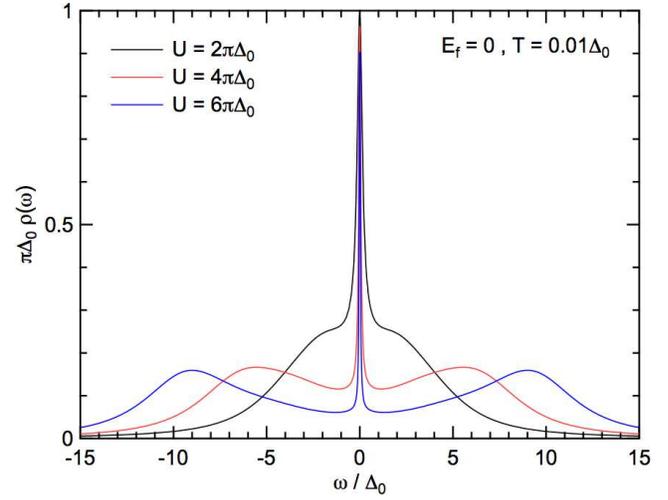}
\end{center}
\caption{(Color online) The $U$ dependence of $\rho(\omega)=-{\rm Im}\,G(\omega)/\pi$ at $T/\Delta_{0}=0.01$.}
\label{dos_U}
\end{figure}

Let us now turn to the one-particle density of states, $\rho(\omega)=-{\rm Im}\,G(\omega)/\pi$.
The $U$ dependence of $\rho(\omega)$ at $T=0.01\Delta_{0}$ is shown in Fig.~\ref{dos_U}.
The central Kondo quasiparticle peak is well developed and its width becomes narrower as $U$ increases.
The position of the Hubbard satellite peaks is slightly smaller than the values of the atomic limit, $\pm U/2$.
The overall features of the DOS are well reproduced and is quantitatively similar to the best know results by the numerical renormalization-group method\cite{Hewson93,Costi94}.

\begin{figure}[tb]
\begin{center}
\includegraphics[width=8.5cm]{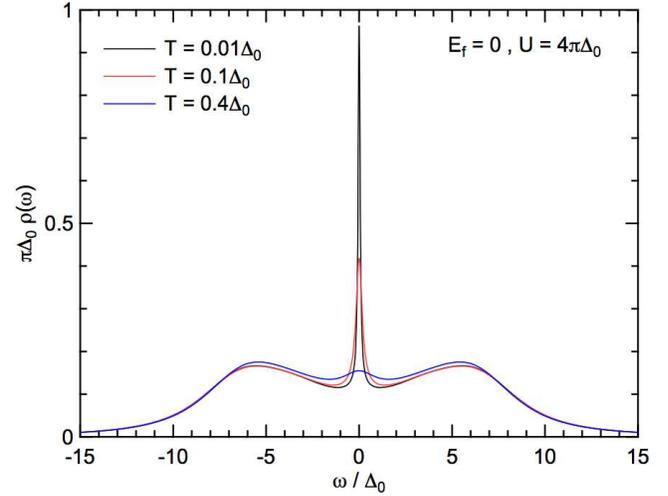}
\end{center}
\caption{(Color online) The $T$ dependence of $\rho(\omega)=-{\rm Im}\,G(\omega)/\pi$ for $U/\pi\Delta_{0}=4$.}
\label{dos_T}
\end{figure}

The $T$ dependence of $\rho(\omega)$ at $U=4\pi\Delta_{0}$ is shown in Fig.~\ref{dos_T}.
The Kondo peak begins to develop below $T_{\rm K}$ where the static spin susceptibility also begins to saturate.
Note that the zero-temperature limit of $\rho(0)$ is fixed as $\rho(0)=1/\pi\Delta_{0}$ due to the Friedel sum rule\cite{Langreth66}.

The quantitative success of the density of states has a common origin of the second-order perturbation theory in the symmetric case\cite{Yosida70,Yamada75,Zlatic85,Horvatic87}.
Namely, the dominant contribution of the fluctuations in (\ref{sefluc}) is the second-order term (the first term in the curly bracket).
By the same reason, the overall features of $\rho(\omega)$ in the asymmetric case (not shown here) are similar to those of the second-order perturbation theory\cite{Zlatic85,Horvatic87}.
Nevertheless, the low-order perturbation theory is inapplicable in higher-dimensional systems in principle, where the phase transitions are possible to occur.
The present approach can describe potential phase transitions in higher dimensions, and at the same time the artificial instabilities of the mean-field theory can be successfully eliminated by the self-consistent treatment of the fluctuations.
The concrete example of the two-dimensional Hubbard model will be shown in the next subsection.

\subsection{Hubbard model on square lattice}
Next, we consider the Hubbard model with the nearest-neighbor hopping,
\begin{equation}
H=\sum_{k\sigma}\epsilon_{k}c_{k\sigma}^{\dagger}c_{k\sigma}
+U\sum_{i}n_{i\uparrow}n_{i\downarrow},
\end{equation}
where $\epsilon_{k}=-2t[\cos(k_{x}a)+\cos(k_{y}a)]$, with $a$ being the lattice constant.
The number operator at $i$ site with the spin $\sigma$ is given by $n_{i\sigma}=c_{i\sigma}^{\dagger}c_{i\sigma}^{}$.
In the following we use $a=t=1$.
In the particle-hole symmetric case at the half filling, we have the relations, $G(k)=-G(k+Q)$, $\Sigma(k)=-\Sigma(k+Q)$, $\chi_{0}(q)=\psi_{0}(q+Q)$ and $\Lambda_{c}=\Lambda_{e}$, where $Q=(\mib{Q};0)$ and $\mib{Q}=(\pi,\pi)$.

\begin{figure}[tb]
\begin{center}
\includegraphics[width=8.5cm]{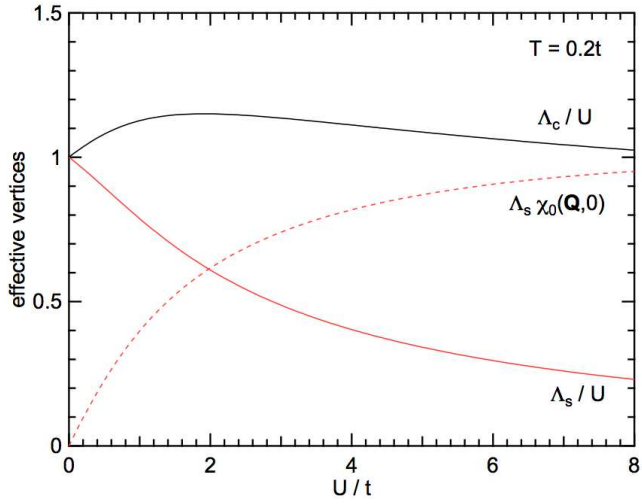}
\end{center}
\caption{(Color online) The $U$ dependence of the effective irreducible vertices for $T=0.2t$.}
\label{U_dep_hub}
\end{figure}

\begin{figure}[tb]
\begin{center}
\includegraphics[width=8.5cm]{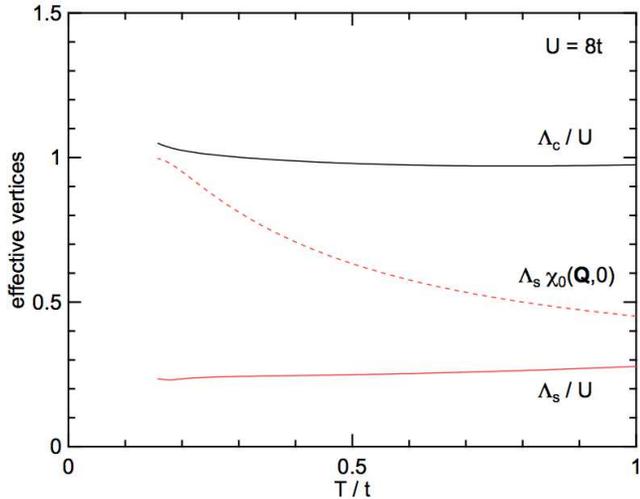}
\end{center}
\caption{(Color online) The $T$ dependence of the effective irreducible vertices for $U=8t$.}
\label{T_dep_hub}
\end{figure}

Let us discuss the symmetric case at the half filling, $n=1/2$.
Figures~\ref{U_dep_hub} and \ref{T_dep_hub} show the $U$ and $T$ dependences of $\Lambda_{c}$ and $\Lambda_{s}$.
The overall behaviors are similar to those of the impurity Anderson model,
{\it e.g.}, $\Lambda_{c}$ is of the order of $U$, $\Lambda_{s}$ is renormalized to eliminate the magnetic instability, and both of $\Lambda_{c}$ and $\Lambda_{s}$ have weak $T$ dependences.

\begin{figure}[tb]
\begin{center}
\includegraphics[width=8.5cm]{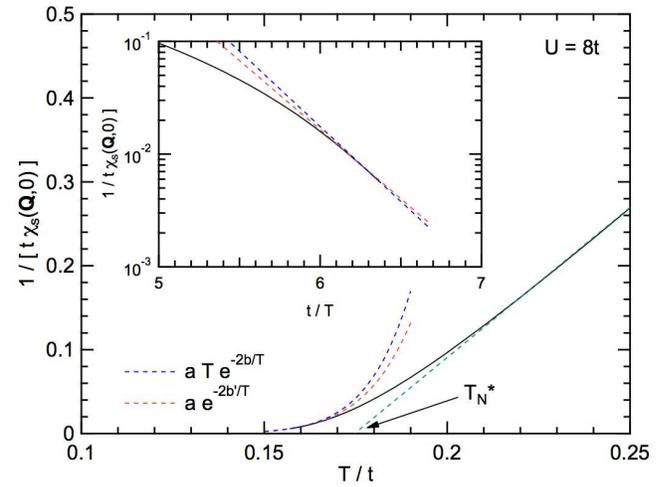}
\end{center}
\caption{(Color online) The $T$ dependence of the inverse static AF spin susceptibility, $1/\chi_{s}(\mib{Q},0)$. The dashed lines are the predicted $T$ dependences $Te^{-2b/T}$ or $e^{-2b'/T}$ with or without the quantum correction. The inset is the logarithmic behavior of $1/\chi_{s}(\mib{Q},0)$ as a function of $1/T$.}
\label{chiSQinv_inset}
\end{figure}

For $T<0.2t$ in Fig.~\ref{T_dep_hub}, $\Lambda_{s}\chi_{0}(\mib{Q},0)$ approaches to unity, but it never exceeds the value of the magnetic instability toward $T=0$.
In order to elucidate the Mermin-Wagner theorem, the $T$ dependence of the inverse staggered susceptibility is shown in Fig.~\ref{chiSQinv_inset}.
$1/\chi_{s}(\mib{Q},0)$ follows the Curie-Weiss behavior in higher temperatures, and it begins to deviate below $T=0.2t$.
For later convenience, we define the fictitious N\'eel temperature, $T_{\rm N}^{*}$ by extrapolating the Cruie-Weiss $T$ dependence.
$T_{\rm N}^{*}$ corresponds to ``mean-field'' N\'eel temperature, and it is regarded as the true critical temperature in the presence of a weak three dimensionality.
The inverse susceptibility is then approaching to zero with the exponential $T$ dependence.
The dashed lines in the figure show the fitting by $T\exp(-2b/T)$ or $\exp(-2b'/T)$ with or without the quantum correction\cite{Moriya90,Vilk97}.
The relation between the present theory and the SCR in the quantum critical region will be discussed in the later section.

\begin{figure}[tb]
\begin{center}
\includegraphics[width=8.5cm]{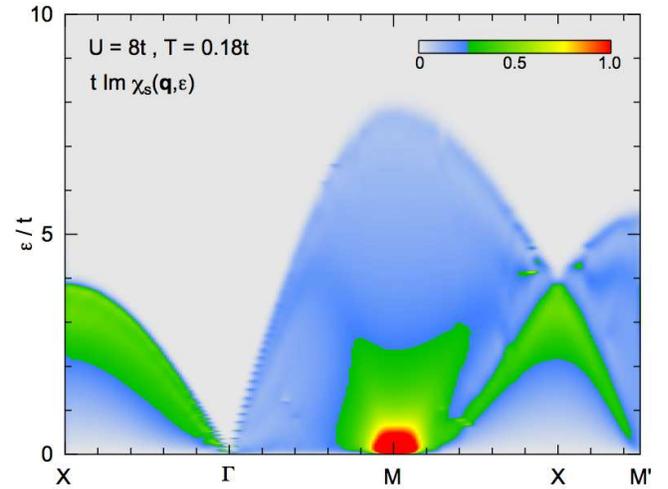}
\end{center}
\caption{The intensity map for ${\rm Im}\,\chi_{s}(\mib{q},\epsilon)$ along the high-symmetry line of the Brillouin zone, where $X=(\pi,0)$, $\Gamma(0,0)$, $M=(\pi,\pi)$ and $M'=(\pi/2,\pi/2)$. The critically enhanced paramagnon excitation is prominent with the sharp peak at $M$ point.}
\label{ImChiS}
\end{figure}

The intensity map of ${\rm Im}\,\chi_{s}(\mib{q},\epsilon)$ for $U=8t$ and $T=0.18t$ along the high-symmetry line of the Brillouin zone is shown in Fig.~\ref{ImChiS}, where $X=(\pi,0)$, $\Gamma=(0,0)$, $M=(\pi,\pi)$ and $M'=(\pi/2,\pi/2)$.
Since the spin susceptibility is of the RPA-type with the renormalized vertex $\Lambda_{s}$, the support of the continuum excitations is given by the possible combinations of the particle-hole pair excitations.
Note that the critically enhanced collective paramagnon excitation is prominent at lower temperatures.

\begin{figure}[tb]
\begin{center}
\includegraphics[width=8.5cm]{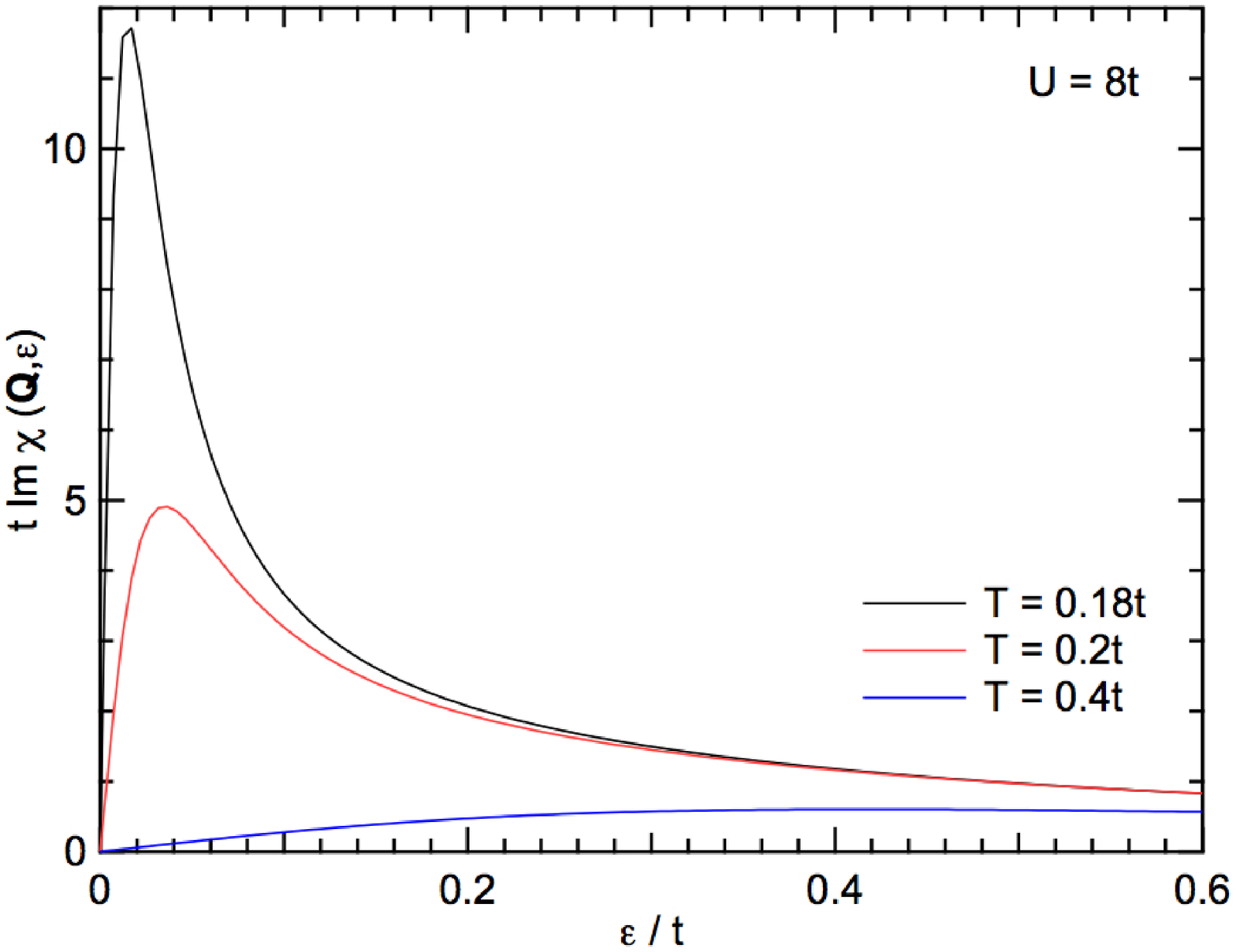}
\end{center}
\caption{(Color online) The $T$ dependence of ${\rm Im}\,\chi_{s}(\mib{Q},\epsilon)$.}
\label{ImChiSQw}
\end{figure}

The imaginary part of the dynamical staggered susceptibility for $U=8t$ is shown in Fig~\ref{ImChiSQw}.
It shows the strong $T$ dependence to enhance the low-energy excitations as $T$ decreases.
As will be shown below, the critical AF spin fluctuations result in the pseudogap behavior in the one-particle excitation, which should give a feedback to the low-energy excitation of ${\rm Im}\,\chi(\mib{Q},\epsilon)$ as well.
Since the present theory lacks the self-consistency between the one-particle and the two-particle levels, no suppressions in the low-energy excitations appears as a feedback.
The full self-consistent treatment remains an open and very challenging problem.

\begin{figure}[tb]
\begin{center}
\includegraphics[width=8.5cm]{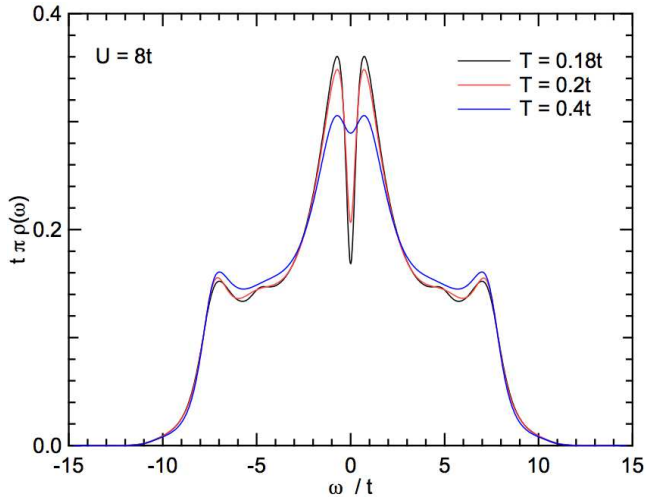}
\end{center}
\caption{(Color online) The $T$ dependence of $\pi\rho(\omega)=-\sum_{\mib{k}}A(\mib{k},\omega)$.}
\label{dosthub}
\end{figure}

Let us now turn to the one-particle properties.
The $T$ dependence of the DOS for $U=8t$ is shown in Fig.~\ref{dosthub}.
As $T$ decreases the central quasiparticle peak develops, and simultaneously the pseudogap in the quasiparticle peak becomes deeper at $\omega=0$.
Note that the development of the pseudogap begins at higher temperature where the staggered spin susceptibility deviates from the Curie-Weiss behavior.
The critical AF spin fluctuations have little influence on the incoherent part, and the transfer of the weight mainly takes place within the quasiparticle peak.
This tendency is in contrast to that of the DMFT with nonlocal corrections where considerable amount of the weight transfers to the incoherent Hubbard satellite peaks from the quasiparticle peak\cite{Maier05,Katanin09,Kusunose06}.

\begin{figure}[tb]
\begin{center}
\includegraphics[width=8.5cm]{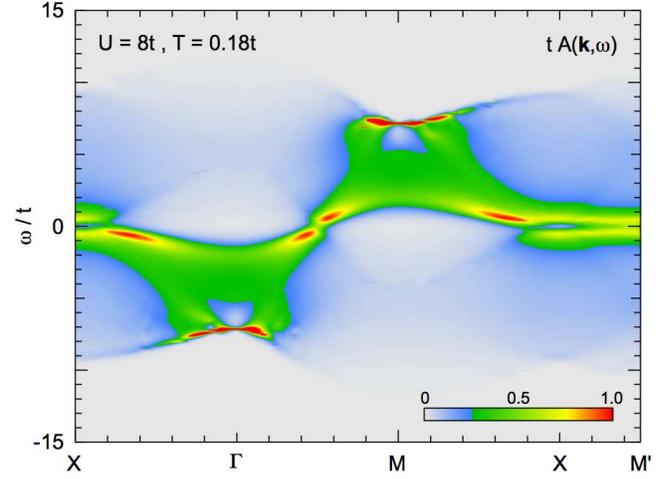}
\end{center}
\caption{(Color online) The $A(\mib{k},\omega)=-{\rm Im}\,G(\mib{k},\omega)$ along the high-symmetry line of the Brillouin zone.}
\label{akw}
\end{figure}

The spectral intensity map of $A(\mib{k},\omega)=-{\rm Im}\,G(\mib{k},\omega)$ along the high-symmetry line of the Brillouin zone for $U=8t$ at $T=0.18t$ is shown in Fig.~\ref{akw}.
Since there is a perfect nesting at half filling and the magnetic Brillouin zone coincides with the Fermi surface, the pseudogap behaviors appear in the coherent quasiparticle bands everywhere on the Fermi surface (see, the X-M' line).
The overall structure of $A(\mib{k},\omega)$ is very similar to that obtained by the DMFT with nonlocal corrections\cite{Katanin09,Kusunose06}.

Now, let us consider the hole doping and the associated superconducting transitions.
In the present theory, the even-pair and the odd-pair irreducible vertices are given by
\begin{align}
&
\Gamma_{e}^{0}(D)=2\gamma+\frac{1}{2}\left[\Phi_{c}(Q)-3\Phi_{s}(Q)\right]+\frac{1}{2}\left[\Phi_{c}(\Omega)-3\Phi_{s}(\Omega)\right],
\cr
&
\Gamma_{o}^{0}(D)=\frac{1}{2}\left[\Phi_{c}(Q)+\Phi_{s}(Q)\right]-\frac{1}{2}\left[\Phi_{c}(\Omega)+\Phi_{s}(\Omega)\right].
\end{align}
Provided the irreducible vertices, we determine the superconducting transition temperature $T_{c}$ by the linearized BS equation,
\begin{equation}
\lambda\phi_{\zeta}(k)=-\frac{1}{2}\sum_{p}\Gamma_{\zeta}^{0}(k,k';0)|G(k')|^{2}\phi_{\zeta}(k'),
\quad
(\zeta=e,o),
\label{approxvpp}
\end{equation}
where $\lambda=1$ signals the superconducting instability of the normal state, and its eigenvector $\phi_{\zeta}(k)$ provides the symmetry of the gap function\cite{Bickers89a}.
Here, $\Phi_{c}(q)$ and $\Phi_{s}(q)$ are given by (\ref{phifluc}).
For the approximate vertices (\ref{approxvpp}), we easily solve the linearized BS equation by using the power method with the fast fourier transformation (FFT) algorithm.
On the Hubbard model, we only consider the $d_{x^{2}-y^{2}}$-type ($B_{1g}$-type) order parameter, which is the favorable symmetry mediated by the AF spin fluctuations.

\begin{figure}[tb]
\begin{center}
\includegraphics[width=8cm]{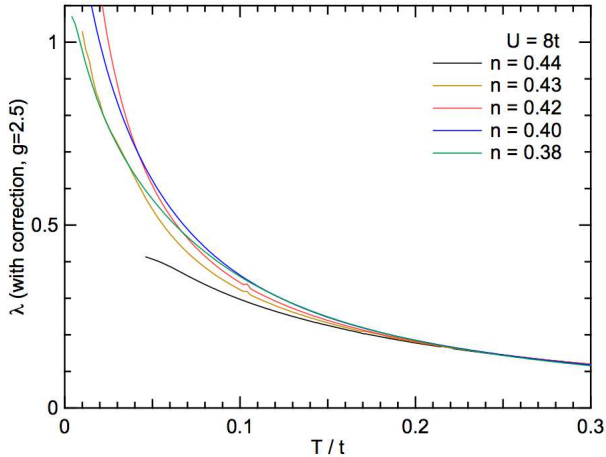}
\end{center}
\caption{(Color online) The $T$ dependence of the eigenvalue of the linearized BS equation.
The $d_{x^{2}-y^{2}}$-type of the order parameter is assumed. In calculating $T_{c}$, the spin vertex is multiplied by a factor $g=2.5$, {\it i.e.}, $\Lambda_{s}\to g\Lambda_{s}$ (see text in detail).}
\label{lamsc1}
\end{figure}

\begin{figure}[tb]
\begin{center}
\includegraphics[width=8cm]{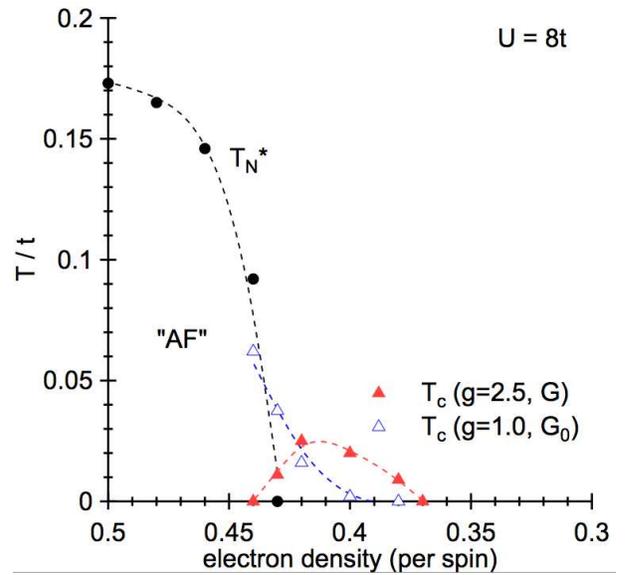}
\end{center}
\caption{(Color online) The $T$-$n$ phase diagram. The $T_{\rm N}^{*}$ is the crossover temperature below which the pseudogap develops. In calculating $T_{c}$, the spin vertex is multiplied by a factor $g$, {\it i.e.}, $\Lambda_{s}\to g\Lambda_{s}$, and using either $G$ or $G_{0}$ (see text in detail).}
\label{phase}
\end{figure}

Away from the half filling, the ``mean-field'' $T_{\rm N}^{*}$ decreases and it vanishes at $n\sim 0.43$ as shown in Fig.~\ref{phase}.
Solving the linearized BS equation, the maximum eigenvalue $\lambda$ does not reach unity at the lowest temperature (not shown) for the doping range, $0.3<n<0.5$.
As compared with the FLEX, which is often used to discuss the superconducting instability, the present theory uses the much smaller coupling constant to the spin fluctuations, $\Lambda_{s}$ renormalized from $U$.
It is the main cause to suppress $T_{c}$ in the present work.
However, the constant approximation for the irreducible vertices may underestimate the coupling constant to the spin fluctuations.
If we considered the momentum and/or the frequency dependence, $\Lambda_{s}(q)$ would have a minimum at $(\mib{Q};0)$ and it would approach to $U$ away from $(\mib{Q};0)$.
Since the strongest fluctuation tends to break the Cooper pairs, the electrons make use of a broad range of fluctuations around the maximum point of $\chi_{s}(q)$ where the coupling constant $\Lambda_{s}(q)$ would be larger than its averaged value.
In order to mimic this effect, we merely replace $\Lambda_{s}$ with $g\Lambda_{s}$ ($g>1$).
The simple enhancement of $\Lambda_{s}$ of course needs to be justified by more elaborate calculations with $q$-dependent vertices.
Nevertheless, it is useful to grasp the doping dependence of $T_{c}$.
For instance, setting $g=2.5$ ($g\Lambda_{s}\sim 0.6U$), we have the $T$ dependence of the eigenvalue as shown in Fig.~\ref{lamsc1}.
In this case, the eigenvalues reach unity with the non-monotonous doping dependence.
Note that $T_{c}$ is suppressed strongly when the system goes into the pseudogap region.
The superconducting phase boundary is indicated by the closed triangle in Fig.~\ref{phase}.
In order to elucidate the retardation effect in the self-energy, we also calculate $T_{c}$ using the bare propagator, {\it i.e.}, $G\to G_{0}$, in the linearized BS kernel (\ref{approxvpp}) with no enhancement factor, $g=1$.
In contrast to the decreasing behavior of $T_{c}$ to the pseudogap region, we obtain the monotonously increasing $T_{c}$ as $n$ increases (the open triangle in Fig.~\ref{phase}).
Therefore, we conclude that the suppression of $T_{c}$ is mainly due to the retardation effect of the self-energy.

\section{Discussions and Summary}

\subsection{Relation to the SCR theory in the vicinity of the QCP}
We discuss the relation between the SCR and the present theory near the QCP.
The core of the SCR theory is the self-consistent equation,
\begin{equation}
\delta(T)=\delta_{0}+g\Braket{\chi_{s}(q)}_{T},
\label{screq}
\end{equation}
where the dynamical spin susceptibility is approximated as
\begin{equation}
\chi_{s}(\mib{Q}^{*}+\mib{q},i\epsilon_{m})^{-1}\sim \delta(T)+A|\mib{q}|^{2}+C\frac{|\epsilon_{m}|}{|\mib{q}|^{z-2}},
\end{equation}
in the critical region, with $z$ being the dynamical exponent.
The $\mib{Q}^{*}$ is a magnetic ordering vector.
The bracket means the $q$-average, $\Braket{\cdots}_{T}=(T/N_{0})\sum_{m}\sum_{\mib{q}}\cdots$, and the subscript $T$ represents that the significant $T$ dependence is included in the average.
The coefficient $g$ is related to the mode-mode coupling constant in the 4th-order free-energy functional.
$\delta(T)=\chi_{s}(\mib{Q}^{*},0)^{-1}$ is a square of the inverse correlation length, and it is  regarded as a measure of the distance from the magnetic QCP.
The peculiar $T$ dependence of $\delta(T)$ determined by (\ref{screq}) governs the critical exponents of various quantities in the QCP region.
Setting $\delta=0$ at $T=0$, we obtain the condition for $\delta_{0}$ to give the QCP,
\begin{equation}
\delta_{0}=-g\Braket{\chi_{s}(q)}_{T=0}.
\end{equation}

In the present theory, if we have the single peak in $\chi_{s}(\mib{q},0)$ at $\mib{q}=\mib{Q}^{*}$, we have the relation,
\begin{equation}
\delta(T)=\chi_{0}(\mib{Q}^{*},0)^{-1}-\Lambda_{s}(T),
\end{equation}
where the weak $T$ dependence in $\chi_{0}$ can be neglected in the QCP region as long as additional singularities such as the van Hove singularity and the perfect nesting are not involved.
The self-consistent equation (\ref{effvcond}) for $\Lambda_{s}$ in the critical region may read
\begin{equation}
\Lambda_{s}=U-c+\alpha\Lambda_{s}^{2}\Braket{\chi_{s}(q)}_{T},
\end{equation}
where $c$ is constant contributions from the charge and the PP fluctuations, and $\alpha$ is an $O(1)$ constant.
In the leading order in $\delta$, it reduces to
\begin{equation}
\delta(T)=\delta_{0}'+g'\Braket{\chi_{s}(q)}_{T},
\end{equation}
which is the same form as (\ref{screq}) with different coefficients.
The formal equivalence of the self-consistent equations thus leads to the same exponent of $\delta(T)$, and other relevant exponents in the QCP region.

Note that in general the basic parquet equations lead to the same critical exponents of a self-consistent $1/N$ expansion for the corresponding order-parameter field theories where $N$ is the number of order-parameter components\cite{Bickers92}.
In contrast to a theory that handles a mode divergence in a single channel, the present theory naturally takes account of interference effects among different modes in the presence of the plural divergences in different channels or multiple characteristic wave vector $\mib{Q}^{*}$.

\subsection{Possible improvements for the present theory}

Let us consider possible improvements for the effective irreducible vertices.
The extension of $\mib{q}$-dependent vertex, $\Lambda(\mib{q})$ is formally straightforward.
Namely, the self-consistent equations (\ref{effvcond}) now become
\begin{equation}
\Lambda_{\xi}(\mib{q})=\gamma+\frac{T\sum_{\epsilon_{m}}\Delta\chi_{\xi}(q)}{T\sum_{\epsilon_{m}}\chi_{0}(q)^{2}},
\quad
\Lambda_{e}(\mib{q})=\gamma+\frac{T\sum_{\epsilon_{m}}\Delta\psi_{e}(q)}{T\sum_{\epsilon_{m}}\psi_{0}(q)^{2}},
\end{equation}
where the $q$ average has been replaced by the $\epsilon_{m}$ average.
Although a formal change is very little, it increases a practical computational effort considerably in the evaluation of the vertex correction terms (\ref{vc}) where the simple FFT algorithm cannot be applied\cite{Ikeda08}.

One may consider that the extension of the full $q$-dependent $\Lambda(q)$ would also be straightforward in a similar way.
However, the self-consistent equations,
\[
\Lambda_{\xi}(q)=\gamma+\frac{\Delta\chi_{\xi}(q)}{\chi_{0}(q)^{2}},
\quad
\Lambda_{e}(q)=\gamma+\frac{\Delta\psi_{e}(q)}{\psi_{0}(q)^{2}},
\]
have an apparent inconsistency.
For example, in the high-frequency limit, the left-hand sides should reduce to the bare interaction, while the right-hand sides diverge as $\epsilon_{m}^{2}$.
This may be due to the fact that only the collective fluctuations do not play a dominant role in the high frequencies, but various individual excitations equally play a role.
All these fluctuations tend to cancel with each other yielding the recovery of the effective irreducible vertices to the bare interaction.
Therefore, a construction of the frequency-dependent vertex theory in this direction seems to be a very challenging problem, unless an appropriate parametrization for the irreducible vertices is found.

\subsection{Summary}

We have proposed the self-consistent fluctuation theory that treats the self-consistency and the antisymmetric nature of the two-particle fluctuations microscopically and appropriately.
The applications to the impurity Anderson model and the Hubbard model on a square lattice have shown the self-consistent treatment of the fluctuations successfully eliminates the finite-temperature magnetic instability of theories without vertex corrections, and it gives the natural energy scale of the spin fluctuations.
In the quantum critical region, the present theory is shown to give the same critical exponents as the SCR theory where a single mode diverges in the spin channel.

The self-energy and the corresponding one-particle spectrum reflect the developed fluctuations in a consistent way via the Schwinger-Dyson equation, and it can be described marginally the Hubbard satellite peaks due to the local spin fluctuations.
Since the DOS at the fermi energy and the double occupancy as a function of the temperature or of the interaction strength do not show a discontinuous change, but show a smooth crossover, the Mott metal-insulator transition cannot be described in the present theory.
Regarding this point, it cannot describe the energy scale of $J\sim t^{2}/U$.
The present theory is thus limited to use in the range of the intermediate coupling, $U\lesssim W$.
To discuss the strong coupling regime, an extension to the frequency-dependent vertices is necessary, and it is a very challenging open problem.

\section*{Acknowledgment}
The author would like to acknowledge stimulating discussions with Hideaki Maebashi, Karsten Held, Andr\'e-Marie Tremblay, V\'aclav Jani\v{s}, Tetsuya Takimoto, Hirokazu Tsunetsugu, Kazuyoshi Yoshimi, Hiroaki Ikeda and Tetsuya Mutou.
He also got benefit from Koichi Izawa who provides inaccessible literatures.
This work was supported by a Grant-in-Aid for Scientific Research on Innovative Areas ``Heavy Electrons" (No. 20102008) of The Ministry of Education, Culture, Sports, Science, and Technology, Japan.

\appendix
\section{Parquet Equation for Single-Band Model with Nonlocal Two-Body Interaction}
\label{appendixa}

In this appendix, we consider the explicit spin dependences of the parquet formalism for single-band models.
For this purpose, we explicitly write the spin indices as $1\to(1\alpha)$, $G(12)\to G_{\alpha\beta}(12)$, $\Gamma(12;34)\to\Gamma_{\alpha\beta;\gamma\delta}(12;34)$ and so on, where the greek letters represent the spin indices.
Let us consider the two-body interaction that conserves the $z$-component of the spins,
\begin{equation}
H_{\rm int}=\frac{1}{2}v_{\alpha\beta}(12)c^{\dagger}_{\alpha}(1)c^{\dagger}_{\beta}(2)c^{}_{\beta}(2)c^{}_{\alpha}(1).
\end{equation}
The anti-symmetrized interactions in the PH and the PP channels are expressed as
\begin{align}
&U_{\alpha\beta;\gamma\delta}(12;34)=v_{\alpha\delta}(14)\left[\delta(12_{+})\delta(3_{+}4)\delta_{\alpha\beta}\delta_{\gamma\delta}
\right.\cr&\hspace{4cm}\left.
-\delta(13_{+})\delta(2_{+}4)\delta_{\alpha\gamma}\delta_{\beta\delta}\right],
\cr
&U^{p}_{\alpha\beta;\gamma\delta}(12;34)=v_{\alpha\beta}(12)\left[\delta(13_{+})\delta(24_{+})\delta_{\alpha\gamma}\delta_{\beta\delta}
\right.\cr&\hspace{4cm}\left.
-\delta(14_{+})\delta(23_{+})\delta_{\alpha\delta}\delta_{\beta\gamma}\right],
\label{antiintp}
\end{align}
where the subscript of the coordinate, e.g. $1_{+}$, specifies the limit of the imaginary time as $\tau_{1}+0$.

When a system conserves the $z$-component of the spins, the one-particle quantities are diagonal in the spin space, {\it e.g.}, $G_{\alpha\beta}=G_{\alpha}\delta_{\alpha\beta}$.
For two-particle quantities, we can decompose the spin dependences into the ``direct'' and ``exchange'' parts only within the spin space as
\begin{align}
&A^{p}_{\alpha\beta;\gamma\delta}(D)=\left[
A^{p(d)}_{\alpha\beta}\delta_{\alpha\gamma}\delta_{\beta\delta}
+A^{p(x)}_{\alpha\beta}\delta_{\alpha\delta}\delta_{\beta\gamma}\right](D),
\cr
&A_{\alpha\beta;\gamma\delta}(D)=\left[A^{(d)}_{\alpha\delta}\delta_{\alpha\beta}\delta_{\gamma\delta}
+A^{(x)}_{\alpha\delta}\delta_{\alpha\gamma}\delta_{\beta\delta}\right](D).
\label{spindep}
\end{align}
Here, $A$ ($A^{p}$) represents any of two-particle quantities such as the full vertex $\Gamma$ ($\Gamma^{p}$), the matrix product $\Phi$ ($\Psi$) and so on.
Note that the bare two-particle green's functions are the special cases with $A^{p(x)}_{\alpha\beta}=0$ and $A^{(d)}_{\alpha\delta}=0$.

The explicit spin dependences of the parquet equation (\ref{parqueteq}) are thus obtained by replacing
\begin{align}
&\Phi(C)\to\Phi_{\alpha\gamma;\beta\delta}(C)=\left[\Phi^{(x)}_{\alpha\delta}\delta_{\alpha\beta}\delta_{\gamma\delta}+\Phi^{(d)}_{\alpha\delta}\delta_{\alpha\gamma}\delta_{\beta\delta}\right](C),
\cr
&\Psi(P)\to\Psi_{\alpha\delta;\gamma\beta}(P)=\left[\Psi^{(x)}_{\alpha\delta}\delta_{\alpha\beta}\delta_{\gamma\delta}+\Psi^{(d)}_{\alpha\delta}\delta_{\alpha\gamma}\delta_{\beta\delta}\right](P),
\cr
&\Phi(X)\to\Phi_{\alpha\gamma;\delta\beta}(X)=\left[\Phi^{(d)}_{\alpha\beta}\delta_{\alpha\gamma}\delta_{\beta\delta}+\Phi^{(x)}_{\alpha\beta}\delta_{\alpha\delta}\delta_{\beta\gamma}\right](X),
\cr
&\Phi(P)\to\Phi_{\alpha\delta;\gamma\beta}(P)=\left[\Phi^{(x)}_{\alpha\beta}\delta_{\alpha\gamma}\delta_{\beta\delta}+\Phi^{(d)}_{\alpha\beta}\delta_{\alpha\delta}\delta_{\beta\gamma}\right](P).
\end{align}
The relevant matrix products are given by
\begin{align}
&\Phi^{(d)}_{\alpha\delta}(D)=-\left[
\Gamma_{\alpha\eta}^{0(d)}\chi_{0,\eta\eta}\Gamma_{\eta\delta}^{(d)}+\Gamma_{\alpha\delta}^{0(d)}\chi_{0,\delta\delta}\Gamma_{\delta\delta}^{(x)}+\Gamma_{\alpha\alpha}^{0(x)}\chi_{0,\alpha\delta}\Gamma_{\alpha\delta}^{(d)}\right](D),
\cr&
\Phi^{(x)}_{\alpha\delta}(D)=-\left[
\Gamma_{\alpha\delta}^{0(x)}\chi_{0,\alpha\delta}\Gamma_{\alpha\delta}^{(x)}\right](D),
\cr
&\Psi^{(d)}_{\alpha\beta}(D)=
-\frac{1}{2}\left[
\Gamma_{\alpha\beta}^{0p(d)}\psi_{0,\alpha\beta}\Gamma_{\alpha\beta}^{p(d)}+\Gamma_{\alpha\beta}^{0p(x)}\psi_{0,\beta\alpha}\Gamma_{\beta\alpha}^{p(x)}\right](D),
\cr&
\Psi^{(x)}_{\alpha\beta}(D)=-\frac{1}{2}\left[
\Gamma_{\alpha\beta}^{0p(d)}\psi_{0,\alpha\beta}\Gamma_{\alpha\beta}^{p(x)}+\Gamma_{\alpha\beta}^{0p(x)}\psi_{0,\beta\alpha}\Gamma_{\beta\alpha}^{p(d)}\right](D).
\label{parquetspin}
\end{align}

Similarly, we obtain the self-energy by using (\ref{antiintp}) as,
\begin{align}
&\Sigma_{{\rm HF}\alpha}(k)=v_{\alpha,-\alpha}(0)n_{-\alpha}-\sum_{q}\left[v_{\alpha\alpha}(q)-v_{\alpha\alpha}(0)\right]G_{\alpha}(k-q),
\cr
&\widetilde{\Sigma}_{\alpha}(k)=-\sum_{pq}v_{\alpha\beta}(k-p)\psi_{0,\alpha\beta}(p;q)\Gamma^{p}_{\alpha\beta;\alpha\beta}(p,k;q)G_{\beta}(q-k),
\end{align}
where $n_{\alpha}=\sum_{k}G_{\alpha}(k)e^{i\omega_{n}0_{+}}$ is the number density for spin $\alpha$.

Now, let us consider the case of SU(2) symmetry in spin space.
In this case, $v_{\alpha\beta}$, $A^{p(d,x)}_{\alpha\beta}$ and $A^{(d,x)}_{\alpha\delta}$ do not depend on the spin indices.
Using the identity,
\begin{align}
\delta_{\alpha\gamma}\delta_{\beta\delta}&
=\frac{1}{2}\left[\left(i\sigma^{y}\right)_{\alpha\beta}^{\dagger}\left(i\sigma^{y}\right)_{\delta\gamma}
+\left(i\sigma^{y}\mib{\sigma}\right)_{\alpha\beta}^{\dagger}\cdot\left(i\sigma^{y}\mib{\sigma}\right)_{\delta\gamma}
\right]
\cr&
=\frac{1}{2}\left[\delta_{\alpha\beta}\delta_{\delta\gamma}
+\mib{\sigma}_{\alpha\beta}\cdot\mib{\sigma}_{\delta\gamma}
\right],
\end{align}
and the transpose property, $(i\sigma^{y}\mib{\sigma})_{12}=(i\sigma^{y}\mib{\sigma})_{21}$, $(i\sigma^{y})_{12}=-(i\sigma^{y})_{21}$, (\ref{spindep}) can be expressed as
\begin{align}
&A^{p}_{\alpha\beta;\gamma\delta}=\frac{1}{2}\left[
A_{e}\left(i\sigma^{y}\right)_{\alpha\beta}^{\dagger}\left(i\sigma^{y}\right)_{\delta\gamma}
+A_{o}\left(i\sigma^{y}\mib{\sigma}\right)_{\alpha\beta}^{\dagger}\cdot\left(i\sigma^{y}\mib{\sigma}\right)_{\delta\gamma}
\right],
\cr
&
A_{\alpha\beta;\gamma\delta}=\frac{1}{2}\left[
A_{c}\delta_{\alpha\beta}\delta_{\delta\gamma}
+A_{s}\mib{\sigma}_{\alpha\beta}\cdot\mib{\sigma}_{\delta\gamma}
\right],
\label{su2int}
\end{align}
where we have defined
\begin{align}
&A_{e}=A^{p(d)}-A^{p(x)}=A^{p}_{\uparrow\downarrow;\uparrow\downarrow}-A^{p}_{\uparrow\downarrow;\downarrow\uparrow},
\cr
&A_{o}=A^{p(d)}+A^{p(x)}=A^{p}_{\uparrow\uparrow;\uparrow\uparrow}=A^{p}_{\uparrow\downarrow;\uparrow\downarrow}+A^{p}_{\uparrow\downarrow;\downarrow\uparrow},
\cr
&A_{c}=2A^{(d)}+A^{(x)}=A_{\uparrow\uparrow;\uparrow\uparrow}+A_{\uparrow\uparrow;\downarrow\downarrow},
\cr
&A_{s}=A^{(x)}=A_{\uparrow\downarrow;\uparrow\downarrow}=A_{\uparrow\uparrow;\uparrow\uparrow}-A_{\uparrow\uparrow;\downarrow\downarrow}.
\end{align}
There is no ambiguity to omit the superscript ``p'' in the case of SU(2) symmetry.
It is obvious from (\ref{su2int}) that $A_{e}$, $A_{o}$, $A_{c}$ and $A_{s}$ represent the two-particle quantities in the even-parity, the odd-parity, the charge and the spin channels, respectively.

Inverting these relations, and substituting them into (\ref{parquetspin}), we obtain,
\begin{align}
&\Phi_{\xi}(D)=-\left[\Gamma^{0}_{\xi}\chi_{0}\Gamma_{\xi}\right](D),
\quad
(\xi=c,s),
\cr
&\Psi_{\zeta}(D)=-\frac{1}{2}\left[\Gamma^{0}_{\zeta}\psi_{0}\Gamma_{\zeta}\right](D),
\quad
(\zeta=e,o).
\end{align}
Similarly, we have the parquet equations in the case of SU(2) symmetry,
\begin{align}
&\Gamma_{c}^{0}(D)=\gamma_{c}(D)-\frac{1}{2}\left[\Phi_{c}+3\Phi_{s}\right](C)
+\frac{1}{2}\left[\Psi_{e}-3\Psi_{o}\right](P),
\cr
&\Gamma^{0}_{s}(D)=\gamma_{s}(D)-\frac{1}{2}\left[\Phi_{c}-\Phi_{s}\right](C)
-\frac{1}{2}\left[\Psi_{e}+\Psi_{o}\right](P),
\cr
&\Gamma_{e}^{0}(D)=\gamma_{e}(D)+\frac{1}{2}\left[\Phi_{c}-3\Phi_{s}\right](X)
+\frac{1}{2}\left[\Phi_{c}-3\Phi_{s}\right](P),
\cr
&\Gamma^{0}_{o}(D)=\gamma_{o}(D)+\frac{1}{2}\left[\Phi_{c}+\Phi_{s}\right](X)
-\frac{1}{2}\left[\Phi_{c}+\Phi_{s}\right](P).
\end{align}
In the basic parquet equations, {\it i.e.}, $\gamma(D)=U(D)$, we obtain
\begin{align}
&\gamma_{c}(D)=2v(q)-v(k-k'),
\cr
&\gamma_{s}(D)=-v(k-k'),
\cr
&\gamma_{e}(D)=v(k-k')+v(k+k'-q),
\cr
&\gamma_{o}(D)=v(k-k')-v(k+k'-q).
\end{align}

The expression of the self-energy in the case of SU(2) is given by
\begin{align}
&\Sigma_{\rm HF}(k)=v(0)n-\sum_{q}\left[v(q)-v(0)\right]G(k-q),
\cr
&\widetilde{\Sigma}(k)=-\frac{1}{2}\sum_{pq}v(k-p)\psi_{0}(p;q)\left[\Gamma_{e}+3\Gamma_{o}\right](p,k;q)G(q-k).
\cr&
\end{align}


\begin{thebibliography}{99}
\bibitem{conservingapprox} The original arguments of the conserving approximation [\citen{Baym62}] aim to give a prescription for the two-particle quantities to be conserving, however, most of recent studies using the FLEX served the conserving self-energy, and discussed mainly for superconducting transitions.
\bibitem{flexmw} The staggered magnetic susceptibility of the FLEX in the expression of the self-energy in the Hubbard model on a square lattice does not show a divergence except at $T=0$. However, this Mermin-Wagner theorem is fulfilled in unphysical way, where the one-particle self-energy is the origin of the suppression of the susceptibility instead of the vertex corrections.
\bibitem{dmftvc} The DMFT and its cluster extensions properly consider vertex corrections in frequency domain, but they neglect or treat insufficiently in the momentum domain.
\bibitem{alprocess} Exactly speaking, the two-particle conservation law requires the Aslamasov-Larkin process with two collective fluctuation exchanges in the particle-hole irreducible vertex, which is not included in the parquet formalism.
\end{thebibliography}
\end{document}